# Impact of System Correlation Matrix Bordering on the MU-MIMO Ergodic Sum-Rate Capacity in the Presence of the Multipath Fading Channel

Aleksey S. Gvozdarev[1], *Member, IEEE*, Aleksandra Alischyuk[2], Marina Kazakova[2]
*Department of infocommunication and radiophysics,*
*P.G. Demidov Yaroslavl State University*
Yaroslavl, Russia
asg.rus@gmail.com[1], yar.radphys@gmail.com[2]

*Abstract*— The research analyzes the problem of capacity analysis of the multi-user massive multi-input multi-output systems with a banded correlation model. The presented study impacts the analytical statistical description with closed form expression of the ergodic sum-rate capacity of the system functioning in presence of multipath fading channel with complex Nakagami-m statistics of the complex transmission coefficients. Numerical verification of the derived expressions was performed and demonstrated excellent correspondence with simulation. The research demonstrated that the system correlation matrix (in case when exponential model is considered) can be bordered up to pentadiagonal structure without introducing any discrepancies in to the sum-rate capacity.

## I. INTRODUCTION

Multiple Input Multiple Output (MIMO) technology is an essential component of most modern wireless communication systems [1]. Moreover, the existing trend on the increase of spectral bandwidth, number of users and supplied data rates extends it to a Multi-User Massive-MIMO modification [2]. It is well known that for such systems, the link quality (measured, for instance, in terms of sum-rate capacity) depends heavily on the possible correlation effects [3] (due to antenna system design) and fading (due to wireless signal multipath propagation) [4]. Both factors impact the so-called MIMO system transmission matrix that accounts for the complex transmission coefficients between any pair of transmitting and receiving antennas.

The performance quality estimation and prediction for such a system rely upon the assumed models of the system correlation matrix [5] and channel model [6]. Moreover, their estimation and update should be performed "on the go" thus, from the practical viewpoint, they should be as simple as possible, retaining the possibility of analytical treatment. However, they should be complex enough to capture most of the propagation and correlation effects. As a classical approach, the correlations within the MIMO channel can be assumed as detached: attributed to the transmitter and receiver separately (Kronecker model) [3], and each correlation matrix (receive and transmit) follow the exponential-type structure [5], [7]. Recently, it was demonstrated that when large antenna arrays are deployed (which is up-to-date for Massive-MIMO) the distant elements can be viewed almost as uncorrelated [8], thus yielding the so-called banded structures [9] allowing us to apply tri- or pentadiagonal iterative algorithms, thus speeding up the calculations.

Concerning the channel model, it should be pointed out that the shrinkage of the coverage area in promising communication systems (for example, 5G, WiFi-7) [10, 11] leads to the discrepancy between the real-life channel measurements and the assumption that the quadrature components of the complex channel transmission coefficients are described by the Gaussian distribution. Hence, some more general models (although losing the potentials for closed-form analytical description) should be applied [4]. One of the possible choices that is considered in this research is the generalization of the well-known Nakagami-m model: complex Nakagami-m distribution [12-13], that operates with instantaneous values rather than with envelope, giving more flexibility in the way the phase statistics is accounted.

Besides, since a multiuser scenario is assumed, we must be careful with the choice of specific user signal detection algorithms since it impacts the sum-rate capacity and computational complexity (computational time) [14]. A standard solution for MIMO systems delivering a compromise between those two factors assumes that the receiver is equipped with the Zero-Forcing (ZF) processing unit, which will be adopted in the research.

The problem of ergodic sum-rate capacity analysis for the case of complex Nakagami-m fading channel has been addressed in literature several times [15]-[19]. It should be noted that the closed-form solutions for most of the MIMO communication problems rely either on the random matrix theory methods (applied to the MIMO system/correlation matrix), on upper/lower bounding or some sort of approximations. The first approach, applied, for instance, in [15], helped to derive the closed-form expression in the cases of 2×2 and 2×3 channel matrices in terms of their joint eigenvalue density functions. Evidently, such an approach fails to deliver closed-form expression for a high-dimensional system due to overcomplication. The second approach (see, for example, [16], [17]) yielded the upper bound for the sum-rate capacity for a system with an arbitrary number of receivers (employing zero-forcing processing) and only 2 transmitters. The third approach (see, for instance, [18], [19]) mostly relies upon the assumption that at some stage of derivations the complex closed-form expressions can be efficiently approximated by some easier ones. A widely accepted proposition (used in [18] for a SIMO system with the maximum ratio combining receiver, in [19] for a MIMO system with the same reception) considers that the signal-to-noise ratio (at the receiver input or before processing, or after postprocessing, etc.) followed gamma/beta/generalized gamma distribution (depending on the case).

The proposed research analyzes the impact of transmitting/receive correlation matrix bordering on overall performance (in terms of the ergodic sum-rate capacity) of the Multi-User Massive-MIMO equipped with zero-forcing processing at the receiving side functioning in the presence of the multipath fading channel with complex Nakagami-m statistics. Moreover, to fill in the gap between the existing results, the closed-form expressions for the sum-rate capacity were derived for arbitrary system size based on the gamma approximation of the ZF postprocessing signal to interference plus noise ratio.

## II. PRELIMINARIES: SYSTEM AND CHANNEL MODELS

### A. General assumptions

Consider that a multi-user MIMO system is communicating in presence of transmit/receive correlation, multipath fading and additive noise. To this extent, we will adopt the following assumptions:

- the correlation is present only on one side (semi-correlated channel model): with the minimum number of antenna elements ($N_T$ at the transmitting side and $N_R$ at the receiver) – the so-called min-semicorrelated model or with the maximum number of antenna elements – the so-called max-semicorrelated model;
- a separable Kronecker model is used to decouple the effects of correlation at the transmitter and receiver, moreover, the correlation matrix structure admitting exponential-type banded model;
- the MIMO multipath complex channel transmission coefficients $\dot{h}_{i,j}$ between any pair of antenna elements on opposite sides follow the complex Nakagami-m distribution;
- the additive noise has Gaussian statistics for the in-phase and quadrature components with equal variances $\sigma^2$, i.e. $z_i \sim \mathcal{CN}(0, \sigma^2)$.

In this case signal vector $\vec{x}$ with the total energy $E_x$ transforms into the received signal vector $\vec{y}$ in the following manner:

$$\vec{y} = \sqrt{\frac{E_x}{N_T}} \begin{cases} \mathbf{\Sigma}_R^{\frac{1}{2}} \mathbf{H}_w \vec{x} + \vec{z}, & N_T > N_R, \\ \mathbf{H}_w \mathbf{\Sigma}_T^{\frac{1}{2}} \vec{x} + \vec{z}, & N_T < N_R, \\ \mathbf{H}_w \mathbf{\Sigma}_T^{\frac{1}{2}} \vec{x} + \vec{z}, & N_T > N_R, \\ \mathbf{\Sigma}_R^{\frac{1}{2}} \mathbf{H}_w \vec{x} + \vec{z}, & N_T < N_R, \end{cases} \quad (1)$$

where the first two cases are applied in the situation of the min-semicorrelation channel and the last two in case of max-semicorrelated, $\mathbf{\Sigma}_R^{\frac{1}{2}}, \mathbf{\Sigma}_T^{\frac{1}{2}}$ are the matrix square roots of the respective correlation matrices, the vector $\vec{z}$ composed of noise samples and $[\mathbf{H_w}]_{i,j} = \dot{h}_{i,j}$ is the fluctuating channel transmission coefficients between any pair of transmitting and receiving antenna elements.

We adopt a classical zero-forcing postprocessing algorithm [4] deployed at the receiver with the output signal to interference plus noise ratio for a $k$-th user is defined as

$$SINR_k = \frac{SNR_{in}}{\left[\left(\mathbf{H}^\dagger \mathbf{H}\right)^{-1}\right]_{k,k}} \quad (2)$$

where $SNR_{in}$ is the average input SNR per stream (receiving antenna element), which is defined relative to $\sigma^2$ (see assumption 4) and either $\mathbf{H} = \mathbf{H}_w \mathbf{\Sigma}_T^{\frac{1}{2}}$ or $\mathbf{H} = \mathbf{\Sigma}_R^{\frac{1}{2}} \mathbf{H}_w$ depending on the side with the existent correlation.

In this case, the primary system performance metric used to quantify the communication link quality is the ergodic sum-rate capacity (ESRC), defined as the expectation of the sum of the capacities for $N$ active users [5]:

$$\bar{C}_\Sigma = \sum_{k=1}^{N} \mathbb{E}\left\{\log_2\left(1 + SINR_k\right)\right\} \quad (3)$$

with $\mathbb{E}\{\bullet\}$ the expectation operator.

### B. Assumption discussions

Commenting on the first assumption (single side correlation), it can be said that in practice it constitutes, for instance, the case of the multi-antenna base station and multiple single-antenna mobile users. In such a case, the correlation side (transmit/receive) depends only on the transmitting mode: uplink or downlink. Since the purely statistical problem formulation is applied (physical differences are omitted) and channel/physical level is under consideration, both situations are valid.

The second assumption naturally arises in the case of large antenna arrays, when the distant antenna elements experience negligibly small mutual correlation (i.e. the correlation coefficient $r_{i,j}$ is small, see, for example, [8]), thus the correlation matrix can be bordered by zeroing its elements with large indices yielding a banded model [9]:

$$[\boldsymbol{\Sigma}_{band}]_{i,j} = \begin{cases} \rho^{|i-j|}, & |i-j| \leq l_{band}, \\ 0, & |i-j| > l_{band}, \end{cases} \quad (4)$$

where $2l_{band}+1$ is the width of the band and $\rho$ is the so-called one-step correlation coefficient [7] that captures the correlation between any pair of adjacent antennas.

The third assumption follows from the need for reasonable generalization of the fading channel model:

- it should be composite, thus incorporating simplified well-studied models as specific limiting cases;
- it should be sufficiently complex to account for most of the propagating effects influencing the channel statistics and be in accordance with the existing real-life measurements;
- it should be simple enough to admit possible analytical treatment.

Thus, for the channel complex transmission coefficients between any pair of transmit-receive antennas, we adopt the complex Nakagami-m model, which was validated to suit real-life scenarios. In this case, the probability density function of the in-phase and quadrature components (i.e. $h_I = \text{Re}(\dot{h}_{i,j})$ or $h_Q = \text{Im}(\dot{h}_{i,j})$) given by [12]:

$$w_h(h) = \frac{m^{\frac{m}{2}} |h|^{m-1}}{\Omega^{\frac{m}{2}} \Gamma(m/2)} \exp\left(-\frac{mh^2}{\Omega}\right), \quad -\infty < h < \infty, \quad (5)$$

where $h = h_I$ or $h = h_Q$, the parameter $\Omega$ stands the average fading power, and $m$ is inversely proportional to the amount of fading.

It is well-known that such a model generalizes several simpler ones: $m=1$ conforms to the Gaussian $h$, thus the Rayleigh envelope, $m=0.5$ corresponds to the one-sided Gaussian envelope model, $m>1$ covers the fading scenarios with the intensities lighter than Rayleigh and $m<1$ correspond to the hyper-Rayleigh model [20-21].

III. DERIVED ANALYTICAL RESULTS

For a model stated in Section II, we proceed with the closed-form derivation of the ESRC (3). First, let us denote the instantaneous $SINR_k$ as $\gamma_k$ and $\xi = \log_2(1+\gamma)$. It is known that in the case when the channel coefficient quadrature components follow the Gaussian model the postprocessing SINR distribution corresponds to the Gamma-type family (including chi-squared). Since the assumed channel model (see Section II) in several particular cases degenerates into those simplified cases we'll assume that $\gamma_k \stackrel{d}{\sim} \Gamma(\hat{\alpha},\hat{\beta})$ (where the notation $\stackrel{d}{\sim}$ stands for "distributed as"). To find the parameter values $\hat{\alpha}, \hat{\beta}$, an intensive numeric simulation was performed. At this stage, for a wide range of system and channel parameters, a large set of samples of $\gamma_k$ were generated and fitted with gamma distribution with the parameters estimated via maximum likelihood procedure in such a way to simultaneously conform with the Pearson $\chi^2$ and the Kolmogorov goodness of fit tests at the statistical significance level 0.05. It was found out that almost for all parameter combinations $\hat{\alpha}_{ML} \approx 1$. Thus for further analysis, we'll assume that $\gamma_k \stackrel{d}{\sim} \Gamma(1, \hat{\beta}_{ML})$.

Applying this assumption and performing the classic random variable transformation procedures

$$w_\xi(y) = w_\gamma(f^{-1}(x)) \cdot \left|\frac{df^{-1}(x)}{dx}\right| = 2^x \ln 2 \cdot \frac{1}{\hat{\beta}} e^{\frac{1}{\hat{\beta}}} e^{-\frac{2x}{\hat{\beta}}}, \quad (6)$$

where $f(x) = \log_2(1+x)$. Thus $\xi \stackrel{d}{\sim} \mathcal{GM}\left(\ln 2, \frac{1}{\hat{\beta}}\right)$, with $\mathcal{GM}(\bullet)$ Gompertz-Makeham distribution [22-23].

Applying the definition of the moment generating function (MGF) $M(\bullet)$ in terms of the Laplace transform, $\mathcal{L}\{\bullet\}$, i.e.

$$M_\xi(s) = \mathcal{L}\{w_\xi(x)\} = \int_0^\infty w_\xi(x) e^{-xs} dx, \quad (7)$$

considering that for the Gompertz-Makeham distribution it is known

$$M_\xi(s) = \hat{\beta}^{-1} e^{\hat{\beta}^{-1}} E_{-\frac{s}{\ln 2}}\left(\hat{\beta}^{-1}\right), \quad (8)$$

where $E(\bullet)$ is the exponential integral [24] and applying the fact that ZF-processing decorrelates $\gamma_k$ the ergodic sum-rate capacity MGF is can be evaluated as:

$$M_{\sum \xi}(s) = \prod_{i=1}^{N} M_{\xi_i}(s) = \frac{e^{\sum_i \frac{1}{\widehat{\beta}_i}}}{\prod_i \widehat{\beta}_i} \prod_{i=1}^{N} E_{-\frac{s}{\ln 2}}\left(\frac{1}{\widehat{\beta}_i}\right) =$$

$$= \frac{e^{\sum_i \frac{1}{\widehat{\beta}_i}}}{\prod_{i=1}^{N} \widehat{\beta}_i} \prod_{i=1}^{N} \left(\frac{1}{\widehat{\beta}_i}\right)^{-\frac{s}{\ln 2}} \widehat{\beta}_i e^{-\frac{1}{\widehat{\beta}_i}} U\left(-\frac{s}{\ln 2}, -\frac{s}{\ln 2}, \frac{1}{\widehat{\beta}_i}\right) = \quad (9)$$

$$= \left(\prod_{i=1}^{N} \widehat{\beta}_i^{\frac{1}{\ln 2}}\right)^{s} \prod_{i=1}^{N} U\left(-\frac{s}{\ln 2}, -\frac{s}{\ln 2}, \frac{1}{\widehat{\beta}_i}\right),$$

where $U(\bullet)$ is the Tricomi confluent hypergeometric function [19].

Applying the identity 13.6.6 [24] helps to derive $M_{\sum \xi}(s)$:

$$M_{\sum \xi}(s) = \left(\prod_{i=1}^{N} \frac{1}{\widehat{\beta}_i}\right) \prod_{i=1}^{N} U\left(1, \; 2+\frac{s}{\ln 2}, \; \frac{1}{\widehat{\beta}_i}\right). \quad (10)$$

The resultant probability density function of the ergodic sum-rate capacity can be equated by inverting the Laplace transform of (10):

$$w_{C_{\sum \xi}}(x) = \mathcal{L}^{-1}\left\{M_{\sum \xi}(s)\right\}. \quad (11)$$

Although (11) cannot be performed analytically there is a wide range of approaches [25] to perform it numerically with high accuracy.

The advantage of the MGF approach is that the average sum-rate capacity can be derived without direct computation of the probability density function. To do this we need to evaluate the first-order derivative of the MGF at $s=0$. Applying the product rule for derivatives yields:

$$\left.\frac{dM_{\sum \xi}(s)}{ds}\right|_{s=0} = \left.\left(\prod_{i=1}^{N} U\left(1, \; 2+\frac{s}{\ln 2}, \; \frac{1}{\widehat{\beta}_i}\right)\right)\right|_{s=0} \times$$

$$\times \left.\left(\sum_{i=1}^{N} \frac{\frac{d}{ds}U\left(1, \; 2+\frac{s}{\ln 2}, \; \frac{1}{\widehat{\beta}_i}\right)}{U\left(1, \; 2+\frac{s}{\ln 2}, \; \frac{1}{\widehat{\beta}_i}\right)}\right)\right|_{s=0} \left(\prod_{i=1}^{N} \frac{1}{\widehat{\beta}_i}\right)$$

$$(12)$$

Applying the identity 13.6.4 from [24] the limiting form of the Tricomi confluent hypergeometric function simplifies to:

$$\left.U\left(1, \; 2+\frac{s}{\ln 2}, \; \frac{1}{\widehat{\beta}_i}\right)\right|_{s=0} = \widehat{\beta}_i, \quad (13)$$

Thus expression for the ergodic sum-rate capacity reduces to:

$$\overline{C}_{\sum \xi} = \left.\frac{dM_{\sum \xi}(s)}{ds}\right|_{s=0} =$$

$$= \left.\left(\sum_{i=1}^{N} \frac{1}{\widehat{\beta}_i} \underbrace{\frac{d}{ds}U\left(1, 2+\frac{s}{\ln 2}, \frac{1}{\widehat{\beta}_i}\right)}_{I_1(s)}\right)\right|_{s=0} \quad (14)$$

The derivative $I_1(s)$ can be evaluated in closed-form:

$$I_1(s) = \frac{\widehat{\beta}_i e^{\frac{1}{\widehat{\beta}_i}}}{\ln 2} G_{2,3}^{3,0}\left(\frac{1}{\beta} \middle| \begin{array}{c} 1, \; 1-\frac{s}{\ln 2}, \; 1-\frac{s}{\ln 2} \\ 1, \; -\frac{s}{\ln 2}, \; -\frac{s}{\ln 2} \end{array}\right) \quad (15)$$

where $G_{2,3}^{3,0}(\bullet)$ is the Meijer G-function [24].

Performing limiting operation

$$\lim_{s \to 0} I_1(s) = \frac{\widehat{\beta}_i e^{\frac{1}{\widehat{\beta}_i}}}{\ln 2} \Gamma\left(0, \frac{1}{\widehat{\beta}_i}\right), \quad (16)$$

the resultant expressions for $\overline{C}_{\sum \xi}$ will be

$$\overline{C}_{\sum \xi} = \frac{1}{\ln 2} \sum_{i=0}^{N} e^{\frac{1}{\widehat{\beta}_i}} \Gamma\left(0, \frac{1}{\widehat{\beta}_i}\right). \quad (17)$$

with $\hat{\beta}_i = \hat{\beta}_{ML_i}$ are the maximum likelihood estimates for each of the $N$ users. Thus the last expression presents the concluding result of the stated problem.

## IV. SIMULATION AND RESULTS

To demonstrate the impact of channel parameters and bordering procedure upon the ergodic sum-rate capacity numeric simulation was performed.

The MIMO system setup under analysis was as follows:

- Multi-User Multiple Input Multiple Output system with an equal number of elements on the transmitting and receiving sides (8 elements), and all users being active ($N=8$).

- Varying fading power $\Omega = \{0.8, 1, 1.2\}$.

- The amount of fading parameter *m* was set to be either 0.7 delivering hyper-Rayleigh, or 2.5 conveying lighter than Rayleigh situation.
- Variable input SNR per receiving antenna: form 0 dB to 20 dB with step 1 dB.
- Varying one-step correlation coefficient $\rho$ from 0 to 0.5.
- Bordering threshold varying from 2 to 8 elements, yielding from tridiagonal to full correlation matrix model.

The results of the simulation are presented in Fig. 1 - Fig. 3.

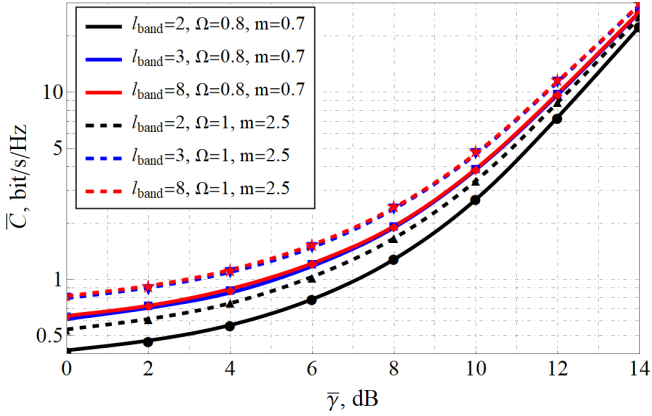

Fig. 1. The ergodic sum-rate capacity of ZF post-processing.

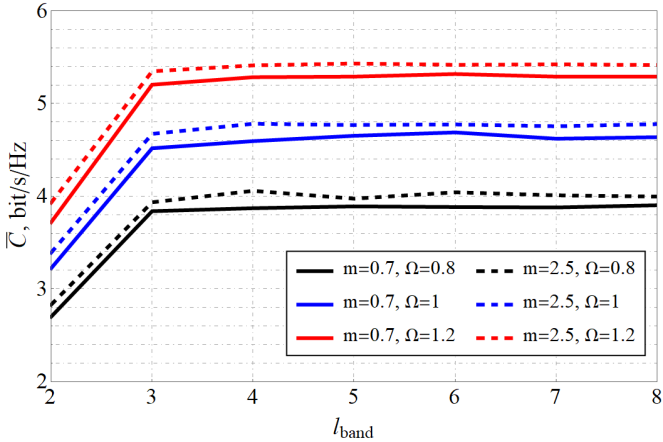

Fig. 2. The ergodic sum-rate capacity of ZF post-processing and per antenna SNR=10 dB.

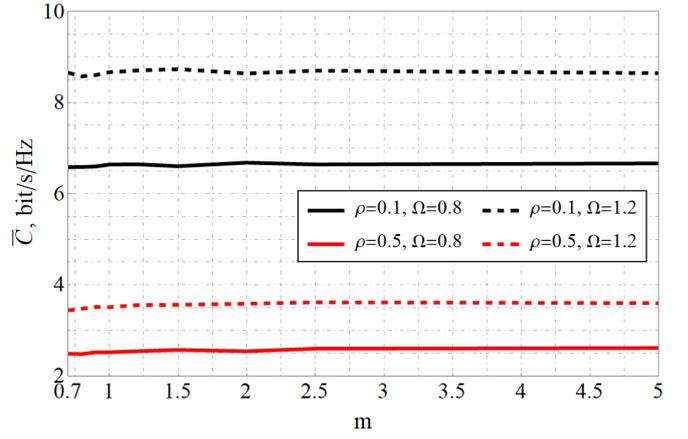

Fig. 3. The ergodic sum-rate capacity of ZF post-processing with $l_{band} = 3$ and per antenna SNR=10 dB.

First, the correctness of the derived approximating ergodic sum-rate capacity expression (17) was verified by comparing it with the numeric simulation.

To do this, a ZF processing (2) for a MU-MIMO system (described by (1), (4) and (5)) was simulated, and then (3) was used to estimate the sum-rate capacity. The results for the two fading scenarios (heavy fading and light fading) are depicted in Fig.1 with markers. The same samples were used to estimate $\hat{\beta}_i$ via maximum likelihood algorithm, and then (17) was used to calculate $\overline{C}_{\sum_\xi}$, predicted by the proposed analytic approximation. It can be seen that the derived solution performs excellently for all the conditions irrespective of the input signal-to-noise ratio.

The carried out analysis demonstrated that for all of the cases the performed post-processing eliminates the impact of the amount of fading parameter.

For all cases under consideration, the improvement of per-antenna SNR (see Fig.1) and reduction of the one-step correlation coefficient (see Fig.3) both lead to the same effect – an increase of ergodic sum-rate capacity.

Analyzing the impact of bordering (see Fig.2), it can be noticed that in the situation of high bordering (tridiagonal correlation matrices) the decrease in the capacity is quite pronounced: up to 1 bit/s/Hz. But starting with $l_{band} = 3$ this decrease diminishes. Hence, high bordering (up to pentadiagonal structures) can be used without significant losses in ergodic sum-rate capacity.

In contrast to banding, the improvement in fading conditions introduces only a minor increase in ESRC (compare solid and dashed lines in Fig.2). Moreover, the increase in the one-step correlation coefficient leads to the smaller impact of channel improvement (compare red and black curves in Fig. 3).

## V. Conclusion

The research studies the ergodic sum-rate capacity of the MU-MIMO system with a one-sided banded model of the system correlation matrix. Under the assumption that at the receiving side zero-forcing processing is employed and that the multipath fading channel transmission coefficients are distributed according to the complex Nakagami-m model, the analytic expression for the sum-rate capacity is derived. The obtained solution was numerically analyzed, and for all of the assumed system parameters demonstrated excellent correspondence with simulation. The research demonstrated that the system correlation matrix (in case when exponential model is considered) can be bordered up to pentadiagonal structure without introducing any discrepancies in to the sum-rate capacity. The results of the research can be of possible interest for the performance prediction of the multiantenna wireless communication systems.